\documentclass[a4paper,fleqn,usenatbib]{mnras}
\usepackage{newtxtext,newtxmath}
\usepackage[T1]{fontenc}
\usepackage{ae,aecompl}
\usepackage{graphics}
\usepackage{graphicx}
\usepackage{amsmath}	
\usepackage[flushleft]{threeparttable}
\usepackage{booktabs,caption}

\title[eRASSU J050810.4-660653]{Spectral and timing analysis of BeXRB eRASSU J050810.4-660653 recently discovered in the Large Magellanic Cloud (LMC)}

\author[Ghising et. al.]{
Manoj Ghising,$^{1}$\thanks{manojghising26@gmail.com}
Mohammed Tobrej,$^{1}$\thanks{tabrez.md565@gmail.com} 
Ruchi Tamang,$^{1}$\thanks{ruchitamang76@gmail.com}
Binay Rai,$^{1}$\thanks{binayrai21@gmail.com}
\newauthor
Bikash Chandra Paul$^{1}$\thanks{bcpaul@associates.iucaa.in}
\\
$^{1}$Department of Physics, North Bengal University,Siliguri, Darjeeling, WB, 734013, India
\\
}
\pubyear{2021}

\begin{document}
\label{firstpage}
\pagerange{\pageref{firstpage}--\pageref{lastpage}}
\maketitle

% Abstract of the paper
\begin{abstract}
 We have studied the Be/X-ray binary (BeXRB) pulsar eRASSU J050810.4-660653 recently discovered in the Large Magellanic Cloud (LMC). Timing and spectral features of the source have been discussed in detail using NuSTAR \& XMM-Newton observations. Coherent pulsation of the source was detected at $\sim 40.578\;\pm\;0.001$ s using NuSTAR observation. We analyzed pulse profiles of the source in different energy bands using NuSTAR \& XMM-Newton data. The pulse-profile evolved with time but was generally  suggestive of a pencil-beam dominated pattern, which combined with the measured luminosity, indicates that the source may be accreting in the sub-critical regime. The pulse fraction follows a linearly increasing trend with photon energy and is anti-correlated with luminosity. In the 1 year interval between the XMM and NuSTAR observations the pulse period shortened by 0.021 s which could be consistent with spin-up or orbital Doppler effect. The average flux of the source in (3-50) keV energy range is found to be $\sim 5.56\;\times\;10^{-11}\;erg\;cm^{-2}\;s^{-1}$ and the corresponding luminosity is $\sim 1.66\;\times\;10^{37}\;erg\;s^{-1}$. The variation of spectral parameters with pulse phase is studied using phase resolved spectroscopy which reveals that the observed photon index becomes harder with increasing flux.
\end{abstract}

%% Select between one and six entries from the list of approved keywords.
%% Don't make up new ones.
\begin{keywords}
accretion, accretion discs-stars: neutron-pulsars: individual: eRASSU J050810.4-660653 – X-rays: binaries.
\end{keywords}

%%%%%%%%%%%%%%%%%%%%%%%%%%%%%%%%%%%%%%%%%%%%%%%%%%%
%
%%%%%%%%%%%%%%%%%% BODY OF PAPER %%%%%%%%%%%%%%%%%%
\section{Introduction}
The Magellanic clouds are known for hosting large number of X-ray binaries particularly High Mass X-ray Binaries (HXMBs). The total number of HMXBs present in the Large Magellanic Cloud (LMC) is relatively lower as compared to Small Magellanic Cloud (SMC). Understanding the evolution of star formation and past histories of LMC \citep{1}, \& SMC \citep{2} can account for the large number of HMXBs that is known today. On the basis of mass of X-ray binaries, they are classified into two categories \textit{viz.} High Mass X-ray Binaries (HXMBs) and Low Mass X-ray Binaries (LMXBs). HMXBs with neutron star (NS) are further categorized into Be/X-ray binaries (BeXRBs) \& Super Giant X-ray Binaries \citep{3}. The optical companion of a BeXRB is a \textit{Be} star which is a non-supergiant fast-rotating B-type belonging to luminosity class III-V stars.  \textit{Be} stars are associated with spectral line emission (primarily the Balmer series) and hence the letter "\textit{e}" in their spectral Type \citep{32}. In addition to their defining hydrogen lines, \textit{Be} stars can also show He \& Fe lines (see e.g \cite{33}). \textit{Be} stars are also known to exhibit extra long-wavelength emission, relative to normal B type stars, this emission is termed as infrared excess. The companion \textit{Be} star in a BeXRB system rotates rapidly which results in an expulsion of photospheric matter along its equatorial plane thereby resulting in the formation of circumstellar disk around it. The circumstellar disk which is also referred as decretion disc \citep{9} evolves continuously \citep{80} allowing the neutron star to capture matter at certain times. The accretion phenomenon is enhanced significantly at periastron passage (when the NS makes its closest approach to the \textit{Be} star) resulting in enhanced X-ray emission by several orders of magnitude for several days. However, when a NS moves away from periastron, accretion from the circumstellar disk ceases which implies that the source returns to the quiescent state. These type of compact objects show transient character and  are studied during bright outbursts by various observatories as the count rate is significantly enhanced. However, Type-II (giant) outburst activity are associated with a higher luminosity (1-2 orders of magnitude more luminous than type I outbursts). Such outbursts are uncorrelated with the orbital phase and  can last longer than one orbital period whereas type I outbursts are connected to the binary orbital phase.  Type-II outbursts are less frequent and hence rarely detected. LXP 38.55 \citep{13}, GX 304-1 \citep{201} are examples of sources  which were reported to undergo type-I outburst while EXO 2030+375 \citep{11}, GRO J2058+42 \citep{202}  are examples of sources associated with type-II outbursts.

 BeXRBs encompass the majority of accreting HMXB pulsars \citep{3}. The typical magnetic fields of HMXB pulsars are of the order of $10^{12}G$ or more \citep{4}. The correlation between the orbital period and the spin period stands out to be one of the significant features of BeXRB systems \citep{5}. Furthermore, the spin period distribution of these class of binary system  follows a bimodal distribution. Such distribution has been proposed as a result of two different types of supernovae viz. iron-core-collapse and electron-capture which produces majority of NS \citep{6}. Other possible linkage between bimodal distribution \& different accretion schemes have been proposed in previous studies \citep{7}. The later proposed schemes have been supported by different long-term X-ray variability of compact objects having both short \& long periods \citep{8}.
BeXRBs may undergo type-I \& type-II outbursts.

The X-ray source eRASSU J050810.4-660653 was discovered in the LMC \citep{14} during the first all-sky survey monitoring by the instrument eROSITA on board Russian/German Spektrum-Roentgen-Gamma (SRG) mission. eROSITA accumulated a total exposure of 1.6 ks. The spectrum provided by eROSITA in the (0.2-8) keV range can be well described by an absorbed power-law with a photon index of $\sim1.2$ \& hydrogen column density of $2.1\times10^{21}\;cm^{-2}$.
The source flux and its corresponding luminosity in the energy range (0.2-10) keV were $3.4\times10^{-12}\;erg\;cm^{-2}\;s^{-1}$ \& $1.2\times10^{36}\;erg\;s^{-1}$, assuming a distance of 50 kpc \citep{14}. Over the seven days of scan of the sky, the background subtracted light curve shows an increasing trend  from 0.24 to 1.14 $ct\;sec^{-1}$ in the energy range (0.2-10) keV. Under the SALT transient followup program, the optical spectroscopy using the Robert Stobie Spectrograph on the Southern African Large Telescope  (SALT)\footnote{\url{https://www.salt.ac.za/}}  further confirmed the source as BeXRB \textbf{\citep{34}}.

The motivation of this paper is to probe  detailed coverage of  spectral \& timing analysis using data from NuSTAR and XMM-Newton observations. The  observation IDs of the source under consideration are 90701342002 (NuSTAR) and 0860800301 (XMM-Newton).

\section{Observation and Data reduction}
The BeXRB eRASSU J050810.4-660653  was observed by Nuclear Spectroscopic Telescope Array (NuSTAR) mission on Dec. 20, 2021. The mission consists of two co-aligned telescopes each having its own focal plane module (FPMA \& FPMB) consisting of a pixelated solid-state CdZnTe detector \citep{15}. Both telescopes FPMA \& FPMB operate in the energy range (3-79) keV. However, due to background contamination above 50 keV, spectral analysis has been carried out in the energy range (3-50) keV. The data reduction was carried out using latest HEASOFT v6.29\footnote{\url{https://heasarc.gsfc.nasa.gov/docs/software/heasoft/download.html}}. Using the mission specific  NUPIPELINE, we created clean event files for analysis. The obtained clean event files are then studied for timing and spectral analysis. The XSELECT tool was used for reading clean event files and hence we extracted counts/sec for FITS light curve, spectra and image. The image  was observed using astronomical imaging and data visualization application DS9\footnote{\url{https://sites.google.com/cfa.harvard.edu/saoimageds9}}. We considered 70 arcsec region around the source for creating source region file. The background region file was also created of the same size by selecting a region far away from the source. Using the above two files, we run the command line NUPRODUCTS for getting necessary light curve FITS file and the spectral PHA file. The background correction for light curve was carried out using FTOOL LCMATH. Barycentric correction was then applied for orbit correction using FTOOL BARYCORR. The spectra  obtained were fitted in XSPEC version 12.12.0.
\begin{table*}
\begin{center}
\begin{tabular}{cllllc}
\hline
Observatory	&	Observation ID	&	Date of Observation (DD-MM-YYYY)	&	Exposure (ks) \\	
\hline
NuSTAR	&	90701342002	&	20-12-2021	&	55.8	\\
XMM-Newton	&	860800301	&	17-12-2020	&	34.2	\\
\hline
\end{tabular}

\caption{NuSTAR \& XMM-Newton observations indicated by their observation IDs along with the date of observation and exposure.}  

\end{center}
\end{table*}
We have also analysed the XMM-Newton archival data observed on Dec.17, 2020. The data processing and extraction were done in XMM-SAS (v20.0.0). Circular source and background regions of 20 arcsec radius respectively were considered for extraction. The events collected by XMM-NEWTON (EPIC-pn detector) provide enough number of counts for period search. The EPIC data were reprocessed using the SAS task EPCHAIN. Event times have been converted to the barycentre of the solar system using the SAS task BARYCEN. We have created a light curve for EPIC-pn with a binning of 0.08 s to estimate the pulse period. 

\subsection{Light curves, pulse profiles and pulse fraction}

\begin{figure*}
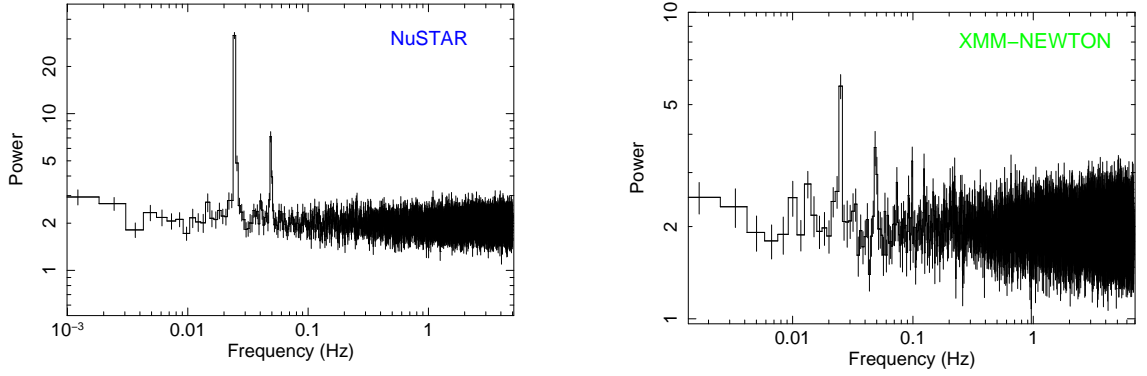


\begin{minipage}{0.3\textwidth}
\includegraphics[height=1.3\columnwidth, angle=-90]{FFT_NuSTAR}
\end{minipage}
\hspace{0.15\linewidth}
\begin{minipage}{0.3\textwidth}
\includegraphics[height=1.3\columnwidth, angle=-90]{FFT_XMM}
\end{minipage}
\caption{FFT of the \emph{NuSTAR} (left) \& \emph{XMM-Newton} (right) light curve of the source revealing the presence of harmonic along with the fundamental signal. The x-axis has been plotted in log scale.}
\label{fig-2}
\end{figure*}

\begin{figure}

\begin{center}
\includegraphics[angle=0,scale=0.3]{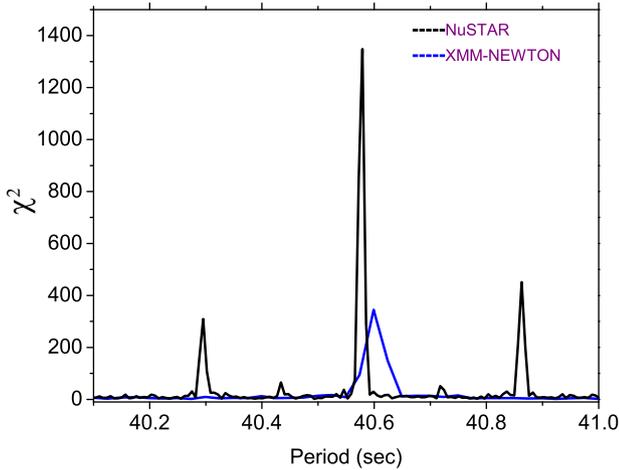}
\end{center}
\caption{Periodogram of the source, for both NuSTAR (black) and XMM-Newton (blue) observation. A coherent signal is clearly detected at $\sim 40.578\;s$ for NuSTAR and $\sim$40.599 s for XMM-Newton.}
\end{figure}

\begin{figure*}

\begin{center}
\includegraphics[angle=0,scale=0.5]{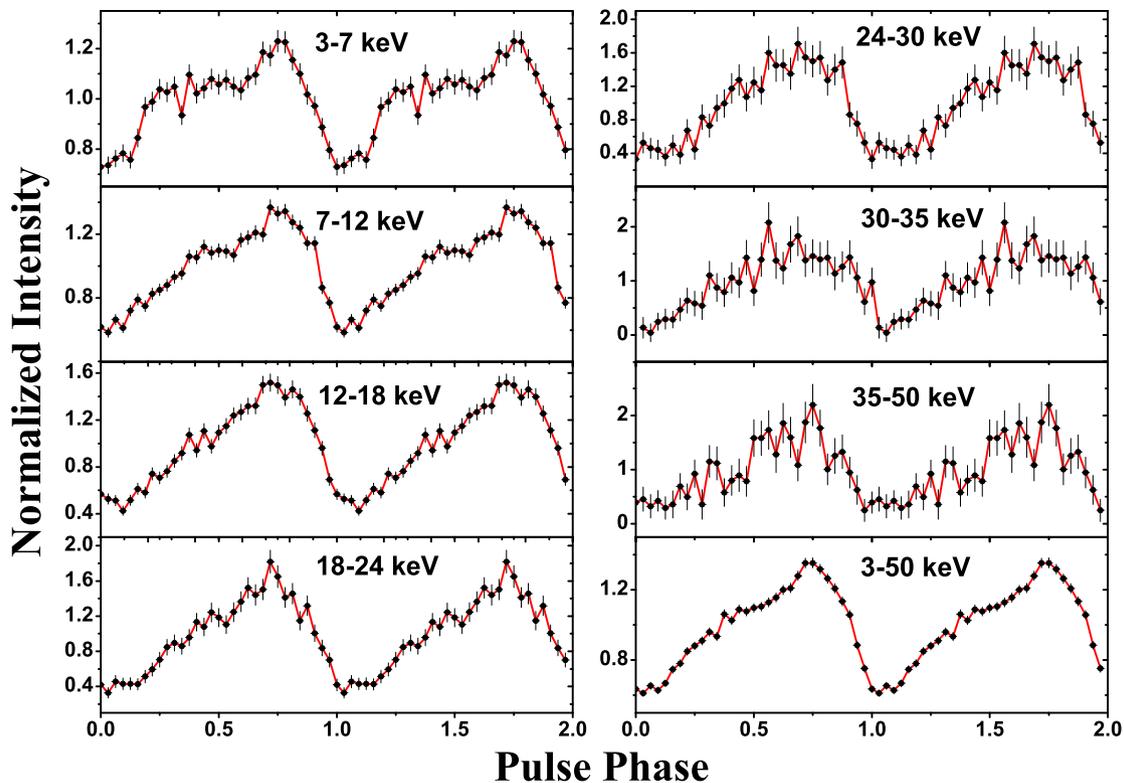}
\end{center}
\caption{ Energy-resolved Pulse profile for NuSTAR observation of BeXRB eRASSU J050810.4- 660653}
\end{figure*}

\begin{figure*}

\begin{center}
\includegraphics[angle=0,scale=0.5]{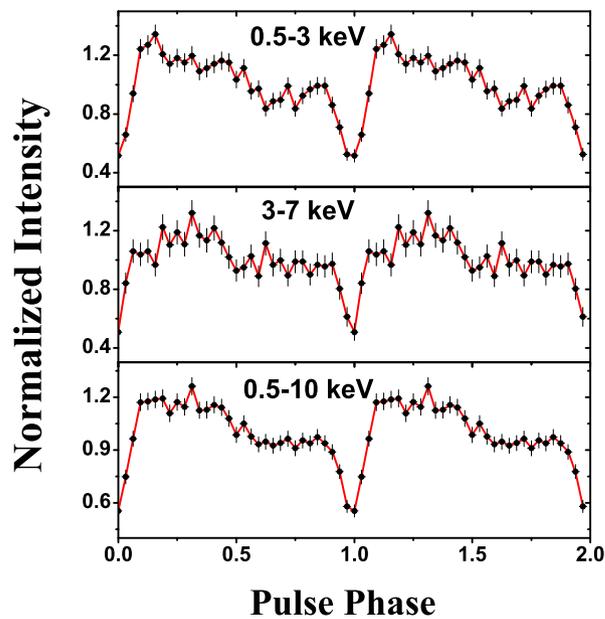}
\end{center}
\caption{Pulse profiles for XMM-Newton observation of BeXRB eRASSU J050810.4-660653 in the low energy band.}
\end{figure*}

\begin{figure}

\begin{center}
\includegraphics[angle=0,scale=0.3]{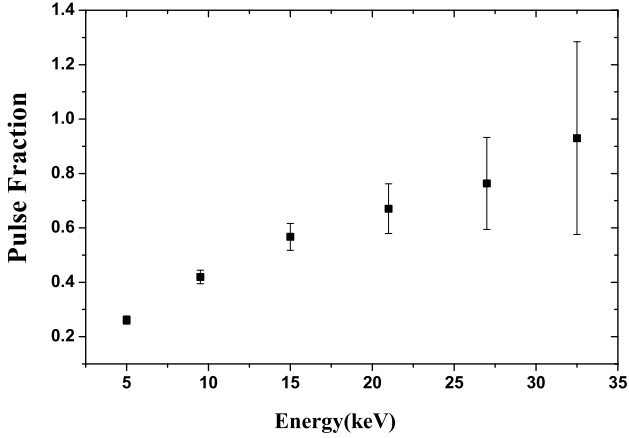}
\end{center}
\caption{Pulse fraction of the source  with energy using NuSTAR observation.}
\end{figure}

\begin{figure}

\begin{center}
\includegraphics[angle=0,scale=0.3]{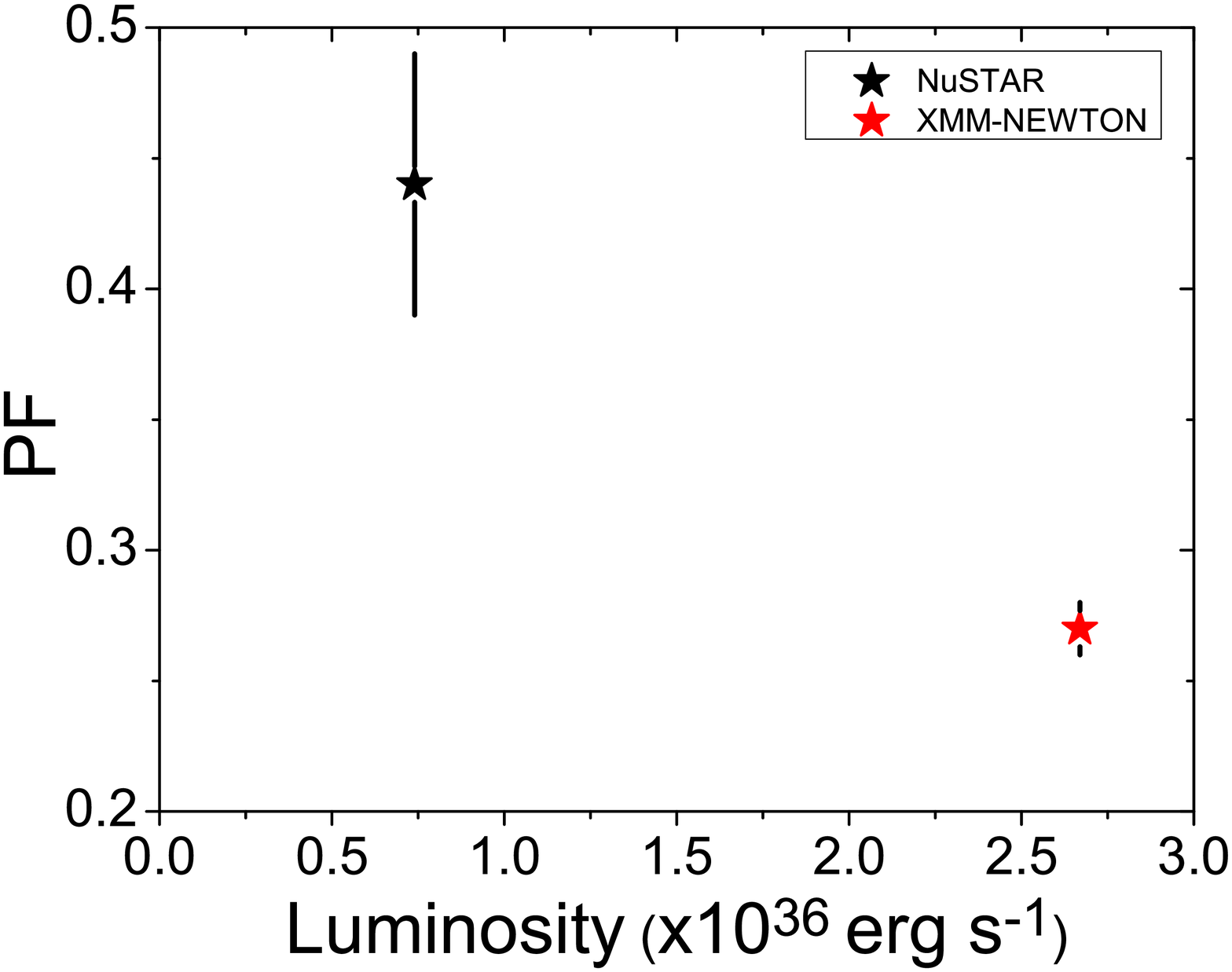}
\end{center}
\caption{Variation of PF with luminosity. NuSTAR \& XMM-Newton data in the common energy range (3-7)keV has been taken into consideration.}
\end{figure}

We consider NuSTAR light curves with binning 0.01 s for analysis of pulse  period and pulse profile. Light curves  were plotted using FTOOL LCURVE. Using Fast Fourier Transform (FFT) of the light curve as shown in Figure 1, the approximate estimation of the pulse period was made. Next, we performed the epoch-folding technique \citep{16}; \citep{17}. This method is based on $\chi^{2}$ maximization technique and hence we estimated the best period at 40.578s as shown in Figure 2. In order to estimate the uncertainty in the measurement of pulsations we employed the method described in \cite{18}. For this purpose, we generated 500 simulated light curves in which the count rate for each time bin is computed using the expression $r_{n}'= r_{n} + \gamma {\sigma_{r}}_{n}$, where  $r_{n}'$ is the count rate of the new light curve at the $n^{th}$  point, $r_{n} $ is the count rate corresponding to the original light curve,  $\gamma$ represents the quantity uniformly distributed in the interval (-1 ; 1) \& ${\sigma_{r}}_{n}$  represents the measurement error of flux for the $n^{th}$ point. We obtained corresponding best period for the derived light curves using the epoch folding technique. We found the best period for each of the 500 samples, then took the mean and standard deviation of these results to arrive at $P_{spin}$ = $40.5786\;\pm\;0.0011\;s$. The FFT of the NuSTAR data revealed a significant harmonic peak at half the fundamental pulse period \textit{i.e.} 20.366 s (see Fig. 1). The XMM-Newton light curve was analyzed in the same way (refer to Fig. 2 which is the periodogram of the XMM data) giving a pulse period $P_{spin}\;= \;40.5997\;\pm\;0.0014\;s$. The FFT and periodogram corresponding to the XMM-Newton observation has been presented in Figure 1 (right) and Figure 2 (blue) respectively. Since these observations as shown in Table 1 were carried out almost exactly one year apart, it appears the pulsar has spun up  by -0.021$\;\pm\;0.002$ \;$s\;yr^{-1}$.

Pulse profiles of the source were then obtained by resolving light curve in the energy range (3-50) keV into several energy bands of (3-7) keV, (7-12) keV, (12-18) keV, (18-24) keV, (24-30) keV, (30-35) keV, \& (35-50) keV. Pulse profiles for NuSTAR and XMM- Newton shown in Figure 3 \& 4 have been folded with a chosen time zero-point $T_0$  (or folding epoch) such that the minimum flux bin lies at phase = 0.0. The NuSTAR pulse profile does not show much variation with energy. In general, the NuSTAR pulse profile is single peaked and it is asymmetric. It has a sawtooth shape at all energies, with the peak emission lying close to phase 0.75. The lowest Nustar energy band (3-7 keV) is different though, it has an additional emission component beginning at phase 0.1. It is difficult to compare the strength of the modulation seen in different energy bands because of the change in count rate.
In order to understand the evolution of pulse profile, we have obtained the pulse profile using XMM-Newton observation as well, the two observations being one year apart. The energy resolved pulse profile were generated in the energy bands (0.5-3) keV, (3-7) keV \& (0.5-10) keV. The XMM pulse contains a single broad pulse, however the sawtooth shape is reversed compared to the NuSTAR profile. XMM sees the peak emission occurring before phase 0.5, in contrast to the NuSTAR profile which peaked after phase 0.5. Perhaps the component seen (only) in the lowest energy NuSTAR profile has increased in strength. We also note that this component is stronger still at the lowest energies (0.5-3 keV) seen by XMM. The evolution of the pulse profiles observed using NuSTAR and XMM-Newton data depicts a changing activity in the accretion geometry. This is also revealed by comparing the common energy resolved  profile (3-7) keV of the system corresponding to NuSTAR \&  XMM-Newton observations.

The Pulse Fraction (PF) is defined as, $PF=\;\frac{P_{max}-P_{min}}{P_{max}+P_{min}}$ where $P_{max} \; \& \; P_{min}$ represents maximum and minimum intensities of pulse profile. The variation of PF is found to be linear revealing a very striking correlation with energy which is shown in Figure 5 which is typical for majority of X-ray pulsars \citep{37}. We have also examined the variation of PF with the luminosity as shown in Figure 6. For this purpose, we have considered the PF of the NuSTAR and XMM-Newton observations in the common energy band (3-7) keV which are associated with different luminosities. Interestingly, the PF was found to be anti-correlated with the luminosity of the source.

\section{Spectral Analysis}
The NuSTAR (FPMA \& FPMB) X-ray spectra of the source eRASSU J050810.4-660653 were fitted in the (3-50) keV energy range which is presented in Figure 7. We ignored the fitting of the source in the energy range above 50 keV due to background count domination. For spectral fitting, we grouped the data using GRPPHA tool to achieve a minimum of 20 counts per spectral bin. The spectra were optimally binned based on the study of \cite{35}.  The relative normalization factors between the two modules were ensured by fixing the constant factor for instrument FPMA as unity without constraining the instrument FPMB to match the average count rate as that in FPMA. This factor corresponding to FPMB  was found out to be 1.02$\;\pm\;0.01$ from the relevant statistical data.  This represents an uncertainty of  $\sim$2\%  which is in good agreement with \cite{20}. The spectral fitting posed a challenge in constraining the $n_{H}$ value. Therefore, we fixed  the absorption column ($n_{H}$) at a value of $1.18\;\times\;10^{22}\;cm^{-2}$  which has been obtained using the HEASARC estimator tool\footnote{\url{https://heasarc.gsfc.nasa.gov/cgi-bin/Tools/w3nh/w3nh.pl}} for the source location in the sky. %(see \cite{21}Ferrigno et al. 2009; cite{22}Iyer et al. 2015; \cite{23}Tsygankov et al. 2016).
 The spectra was then fitted by using a single continuum model CONSTANT*PHABS*CUTOFFPL. The spectral fitting with the given model estimates the cutoff energy at 20.10 keV \& photon index at 0.84. The applied model reduces the fit statistics i.e $\chi^{2}$ per degrees of freedom to about $\sim$ 1.04. Though weak residuals are seen in the fitted spectra, however, these residuals are insignificant and do not contribute to emission or absorption components. The average flux was then calculated in the energy range (3-50) keV. The estimated value of the flux turns out to be around $\sim\;5.56\;\times\;10^{-11} erg\;cm^{-2}\;s^{-1}$ and hence the luminosity being $\sim\;1.66\;\times\;10^{37} erg\;s^{-1}$ assuming the source to be at a distance of 50 kpc \citep{14}.

In the low energy range, the spectra was suitably fitted by considering the XMM-Newton observation using the absorbed power-law model. The single continuum model CONSTANT*PHABS*POWERLAW generated the best fit results in the soft energy band (0.5-10) keV. The fitted spectra has been presented in Figure 8 and the corresponding statistical data has been presented in Table 2. The absorbed flux of the system in the said energy range was found to be \;$\sim\;5.62\;\times\;10^{-12}\;erg\;cm^{-2}\;s^{-1}$ which is equivalent to a luminosity of $\sim1.7\;\times\;10^{36}\;erg\;s^{-1}$. 

\begin{figure}
\begin{center}
\includegraphics[angle=270,scale=0.3]{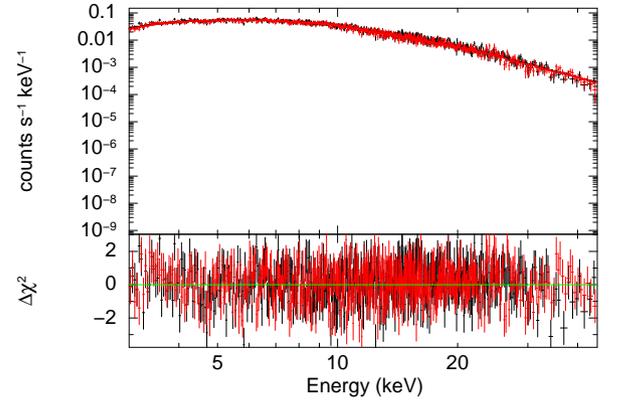}
\end{center}
\caption{The folded spectra of NuSTAR observation using the continuum model CONSTANT*PHABS*CUTOFFPL. The bottom panel shows residuals. Black \& red spectra represent NuSTAR FPMA \& FPMB spectra respectively.}
\end{figure}
\begin{figure}

\begin{center}
\includegraphics[angle=270,scale=0.3]{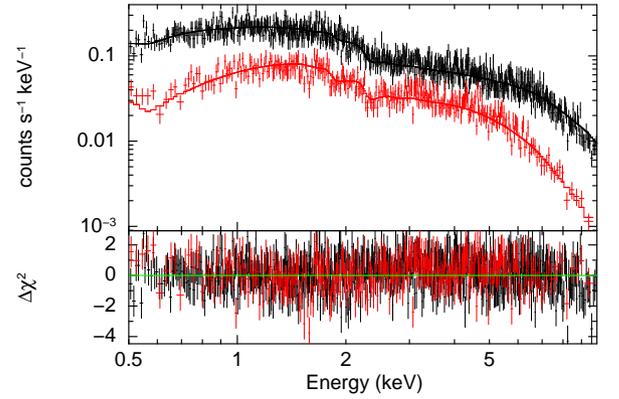}
\end{center}
\caption{The folded spectra of XMM-Newton observation using the continuum model CONSTANT*PHABS*POWERLAW. The bottom panel shows residuals. Black \& red spectra represent  PN \& MOS1 spectra respectively.}
\end{figure}

\begin{table*}
 \begin{center}
 \begin{tabular}{clc}
    \hline
   Spectral parameters	&	Data (NuSTAR)	&	Data (XMM-Newton)	\\

\hline					
$C_{FPMA}$	&	1.0 (fixed)	&	\dots	\\
$C_{FPMB}$	&	$1.02\;\pm\;0.01$	&	\dots	\\
$C_{PN}$	&	\dots	&	1.0 (fixed)	\\
$C_{MOS1}$	&	\dots	&	1.04 $\pm$ 0.02	\\
$n_{H}\;(cm^{-2})$	&	1.18 (fixed)	&	1.18 (fixed)	\\
$\alpha$	&	0.84$ \;\pm\;0.03$	&	0.82$\;\pm\;$0.01	\\
$\alpha_{norm}\;(\times\;10^{-4}\;ph\;cm^{-2}\;s^{-1})$	&	\dots	&	$2.81\;\pm\;0.1$	\\
$E_{C}$ (keV)	&	20.10$\;\pm\;$0.95	&	\dots	\\
$norm_{cutoffpl}\; (\times\;10^{-3} ph\;cm^{-2}\;s^{-1}) $	&	1.45$\;\pm\;$0.1	&	\dots	\\
	&		&		\\
$\chi^{2}$	&	1030.65	&	1023.07	\\
Degrees of freedom	&	982	&	992	\\
$\chi_{\nu}^{2}$	&	1.05	&	1.03	\\
Flux (erg $cm^{-2}\;s^{-1}$)	&	(5.56$\;\pm\;0.1)\;\times\;10^{-11}$	&	(5.62$\;\pm\;0.06) \;\times\;10^{-12}$	\\

   \hline
  \end{tabular}
  \caption{The above table lists the fit parameters of eRASSU J050810.4-660653 using model combination CONSTANT*PHABS*CUTOFFPL for NuSTAR observation and CONSTANT*PHABS*POWERLAW for XMM-Newton observation. Photon Index ($\alpha$) \& CUTOFF energy ($E_{C}$) are the spectral parameters of the CUTOFFPL model. Flux was calculated within energy range (3-50) keV for NuSTAR and  within energy range (0.5-10) keV for XMM-Newton. $\chi_{\nu}^{2}$ represents reduced $\chi^{2}$. Errors quoted for each parameters are within 90\% confidence interval.}
  \end{center}
 \end{table*}

\subsection{Phase Resolved Spectral Analysis}
Phase resolved analysis was carried out for understanding the variation of spectral parameters with pulse phase. For this purpose, we have divided the pulse period into 10 segments. For each segment, we created good time interval (GTI) files using tool XSELECT and then merged them with tool MGTIME.  After successfully creating merged GTI files, the required spectra files were generated using the NUPRODUCTS package.

The spectra corresponding to all the phase bins have been fitted using the  continuum model CONSTANT*PHABS*CUTOFFPL. The hydrogen column density in phase resolved spectral study has been fixed at $1.18\;\times\;10^{22} cm^{-2}$. For each phase bin, flux has been estimated in the (3-50) keV energy range. The nature of flux variation is analogous to the pulse profile. The flux varied between the peak value of $7.95\;\times\;10^{-11} erg\;cm^{-2}\;s^{-1}$ in the phase interval (0.6-0.7) \& (0.7-0.8) and minimum value of $2.68\;\times\;10^{-11}\;erg\;cm^{-2}\;s^{-1}$ in the phase interval between (0.9-1). The variation of photon index with phase also varies between maximum value of 1.16 in the phase interval (0.1-0.2) and minimum value of 0.67 in the phase interval (0.6-0.7) \& (0.7-0.8). The minimum value of $E_{C}$ with phase lies in the phase interval (0-0.1) at 10.92 keV while the maximum value corresponds to phase interval (0.3-0.4) at 31.02 keV. The flux is anti-correlated with the power-law photon index which is evident from phase-resolved spectral study  represented in Figure 9.

\begin{figure}

\includegraphics[angle=0,scale=0.35]{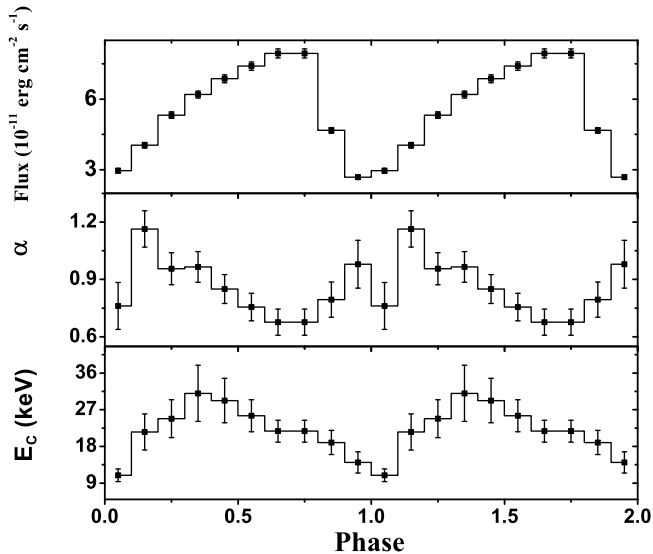}

 \caption{Variation of spectral parameters: Flux, Photon Index ($\alpha$) \& CUTOFF Energy ($ E_{C}$) with phase for NuSTAR observation of eRASSU J050810.4-660653.}
\end{figure}

\section{Discussion \& Conclusion}
The available {NuSTAR and XMM-Newton data for the source} have been used to analyze timing and spectral properties of this BeXRB system. Timing analysis of NuSTAR data detected the coherent pulsations at $40.5786\;\pm\;0.0011$ s. The same analysis of XMM-Newton data detected the pulsation to be $40.5997\;\pm\;0.0014$ s. Since, these observations have been made at different spans of time, it appears that the pulsar has spun-up by -0.021 s\;$yr^{-1}$. However, we are unaware of the effects of orbital Doppler shift on spin period as the source was discovered recently. Such effects may be explored in the long run with the availability of more observational data.

Transformation of pulse profiles from multi-peaks to single peak from lower energies to higher energies are commonly observed in many pulsars \citep{19}. Such features have been observed in Her X-1 \citep{26}, 4U 0115+63 and V 0332+63. This type of variations are typically observed in various X-ray pulsars especially bright ones (\cite{37}). A qualitative interpretation of pulse-profile dependency has been proposed based on the results of \citep{27}. The variation of pulse profile of X-ray pulsar V 0332+63  was interpreted in a purely geometric fashion. Among accretion-powered pulsars, pulse profile of BeXRBs are found to be complex during outbursts. The pulse profile in our timing analysis is characterised by single peaked feature throughout the observations. The observed pulse profiles corresponding to the source are  asymmetric in nature which are typically observed in X-ray pulsars. 
Various theoretical models have been formulated to explain the asymmetric nature  of the pulse profiles. The distorted magnetic dipole field, where the  magnetic poles are not exactly opposite to one another other, may be a probable reason for the asymmetric nature of pulse profiles (\cite{103} ; \cite{104} ; \cite{105}; \cite{106}). An asymmetric accretion
stream has been suggested as another possible reason for the asymmetry of pulse profiles (\cite{100}; \cite{101}; \cite{102}). As evident from the pulse profile morphology presented in Figure 3 \& 4, the profiles are not sinusoidal in nature which gives  an insight about the pattern of harmonics revealed in the power-spectrum of the source (see Fig. 1). The pulse profile is found to evolve with time thereby providing a conclusive evidence of changing accretion geometry. This fact can simply be visualised from the variations in the pulse profile observed in the two sets of data observed at different spans of time. The single peaked feature reflects a pencil beam pattern in the NS radiation. A pencil-beam pattern is also characterized by luminosities lower than the critical luminosity ($L_{c})$ of the source. $L_{c}$  marks the transition in the accretion regime of the source. If the luminosity of the source is lower than $L_{c}$, the accretion phenomena is known to occur in the sub-critical regime while for luminosities greater than $L_{c}$, the accretion phenomena occurs in the super-critical regime of the pulsar. Hence, the pencil-beamed pattern reveals the fact that the source may be accreting in the subcritical regime. In the case of a ‘pencil beam’ pattern, the accreted material reaches the surface of the NS through nuclear collisions with atmospheric protons or through coulomb collisions with thermal electrons \citep{81}. The emission phenomena take place from the top of the column \citep{82}. Therefore, it is appropriate to express that the X-ray luminosity of the source is lower than the critical luminosity. The luminosity dependence of the pulse profiles reflects an alteration in the physical state of the accretion column. In case of accretion in the sub-critical regime, the height of the accretion column is known to be anti-correlated with the accretion rate which may be a probable reason for the luminosity dependence of the pulse profiles \citep{62}.
 
  The smoothness of the PF plot gives another interesting characteristic of the source. This is consistent with the absence of strong features such as CSRF in the energy spectrum of the source. The non-monotonic dependence of the PF on energy around the cyclotron feature has been reported for various X-ray pulsars (\cite{38}; \cite{37}; \cite{39}).
On examining the variation of PF with luminosity, it was observed that the PF follows an anti-correlated pattern with luminosity. This may have arised due to an increment in the unpulsed component of the total emission or a net decrement of the pulsed component\citep{85}.

The NuSTAR spectrum was best-fitted by an absorbed CUTOFFPL model revealing a photon index of 0.84 \& cutoff energy of 20.10 keV. The average flux of the source in (3-50) keV energy range was found to be $\sim 5.56\;\times\;10^{-11}\;erg\;cm^{-2}\;s^{-1}$ which corresponds to a luminosity of $\sim 1.66\;\times\;10^{37}\;erg\;s^{-1}$ assuming a distance of 50 kpc. Usually, residuals are seen in the spectral fit of many BeXRB systems when fitted with power-law, therefore an additional component like soft blackbody is  generally required to obtain the best fit statistics (e.g. \cite{21}; \cite{22}; \cite{23}; \cite{24} ). However, black body model was not required as the spectra was suitably fitted without its implementation as positive residuals in the soft energy band were not prominent. Various BeXRBs like 1A 0535+26, V 0332+53, Cep X-4, 4U 0115+63 and many more are known to exhibit significant CRSF feature (\cite{42}). However, such characteristic features were not observed in the source. We do not find the presence of fluorescent lines of iron  in the observed spectra. The absence of additional spectral features were also verified by phase resolved spectroscopy, as various sources are known in which the cyclotron line or its higher harmonics are prominent only at certain phases of rotation (e.g \cite{41}). It is evident from Figure 9 that the cut-off energy varies with pulse phases. The cut-off energy has no clear physical meaning although it may be related to the strength of the magnetic field \citep{110}. An empirical relation between exponential cut-off energy and cyclotron line energy has been established by \cite{110} which is beyond the scope of our study for this source. However, it has been observed that for cut-off energies in the range (8-12) keV, the cyclotron line energies are inferred to lie in the range (12-40) keV \citep{110}. In the present study of the source, the cut-off energy is found to be about 20 keV which suggests that the cyclotron energy (if present) must be at higher energies (> 50 keV). Since, we have ignored the energy spectrum above 50 keV due to background domination, the absorption feature (if present) may be detected with the availability of more observational data.
 Interestingly, the power-law photon index was found to vary with the pulse phase of the system. It is evident from Figure 9 that when the source flux is high, the system becomes harder and when the source flux is low, the system becomes softer. The increase in luminosity is associated with an increase in optical depth. This in turn leads to an increase in the hardness of photons which is in agreement with hardening (decrease in photon index) of the spectrum with increase in flux of the system \citep{40}.

 \section*{Acknowledgements}
 This research work has been carried out by using publicly available data of the pulsar provided by NASA HEASARC data archive. The NuSTAR data used in the research is provided by NASA High Energy Astrophysics Science Archive Research Center (HEASARC), Goddard Space Flight Center. We further acknowledge the use of public data from XMM-Newton observatory. We would like to thank the anonymous reviewer for his/her valuable suggestions to improve the quality of the paper. We are grateful to IUCAA Centre for Astronomy Research and Development (ICARD), Department of Physics, NBU, for providing research facilities.  
 
 \section*{Data availability}
 
 The observational data used in this study can be accessed from the HEASARC data archive and is publicly  availaible for carrying out research work.

%
%
%
%

%
%

%%%%%%%%%%%%%%%%%%%%%%%%%%%%%%%%%%%%%%%%%%%%%%%%%%

%%%%%%%%%%%%%%%%%%%% REFERENCES %%%%%%%%%%%%%%%%%%

% The best way to enter references is to use BibTeX:

%\bibliographystyle{mnras}
%\bibliography{example} % if your bibtex file is called example.bib

% Alternatively you could enter them by hand, like this:
% This method is tedious and prone to error if you have lots of references

%%%%%%%%%%%%%%%%%%%%%%%%%%%%%%%%%%%%%%%%%%%%%%%%%%

%%%%%%%%%%%%%%%%% APPENDICES %%%%%%%%%%%%%%%%%%%%%

%%%%%%%%%%%%%%%%%%%%%%%%%%%%%%%%%%%%%%%%%%%%%%%%%%

% Don't change these lines
\bsp	% typesetting comment
\label{lastpage}
\end{document}